\documentclass[a4paper,11pt]{article}
\usepackage{pos}
\usepackage{cleveref}
\usepackage{tikz}
\usetikzlibrary{matrix,decorations.pathreplacing,calc,fit,backgrounds}
\usepackage{dsfont}
\usepackage{bm}
\usepackage{mathtools}

\usepackage[skins,theorems]{tcolorbox}
\tcbset{highlight math style={enhanced,
  colframe=red,colback=white,arc=0pt,boxrule=1pt}}

\newtcolorbox{mybox}[2][]{enhanced,
attach boxed title to top left={xshift=5mm,yshift=-2mm},
  title={#2},#1}

\newcommand{\td}{\dot t}
\newcommand{\tp}{t^\prime}

\newcommand{\pd}{\dot \phi}
\newcommand{\pp}{\phi^\prime}

\title{Exact space-time symmetry conservation and automatic mesh refinement for classical lattice field theory}
\ShortTitle{Symmetry preservation and mesh refinement in classical LFT}

\author*[a]{A.~Rothkopf}
\author[b]{W.~A.~Horowitz}
\author[c,d]{J.~Nordstr\"om}

\affiliation[a]{Faculty of Science and Technology, University of Stavanger, 4021 Stavanger, Norway}

\affiliation[b]{Department of Physics, University of Cape Town, 7701 Rondebosch, South Africa}

\affiliation[c]{Department of Mathematics, Link{\"o}ping University, SE-581 83 Link{\"o}ping, Sweden}

\affiliation[d]{Department of Mathematics and Applied Mathematics, University of Johannesburg, P.O. Box 524, Auckland Park 2006, Johannesburg, South Africa}

\emailAdd{alexander.rothkopf@uis.no}

\abstract{The breaking of space-time symmetries and the non-conservation of the associated Noether charges constitutes a central artifact in lattice field theory. In \cite{Rothkopf:2023ljz,Rothkopf:2023vki} we have shown how to overcome this limitation for classical actions describing point particle motion, using the world-line formalism of general relativity. The key is to treat coordinate maps (from an abstract parameter space into space-time) as dynamical and dependent degrees of freedom, which remain continuous after discretization of the underlying parameter space. Here we present latest results \cite{Rothkopf:2024hxi} where we construct a reparameterization invariant classical action for scalar fields, which features dynamical coordinate maps. We highlight the following achievements of our approach: 1) global space-time symmetries remain intact after discretization and the associated Noether charges remain exactly preserved 2) coordinate maps adapt to the dynamics of the scalar field leading to adaptive grid resolution guided by the symmetries.}

\FullConference{The 41st International Symposium on Lattice Field Theory (LATTICE2024)\\
 28 July - 3 August 2024\\
Liverpool, UK\\}

\begin{document}
\maketitle

\section{Motivation}

\begin{figure}
\centering
\includegraphics[scale=0.43]{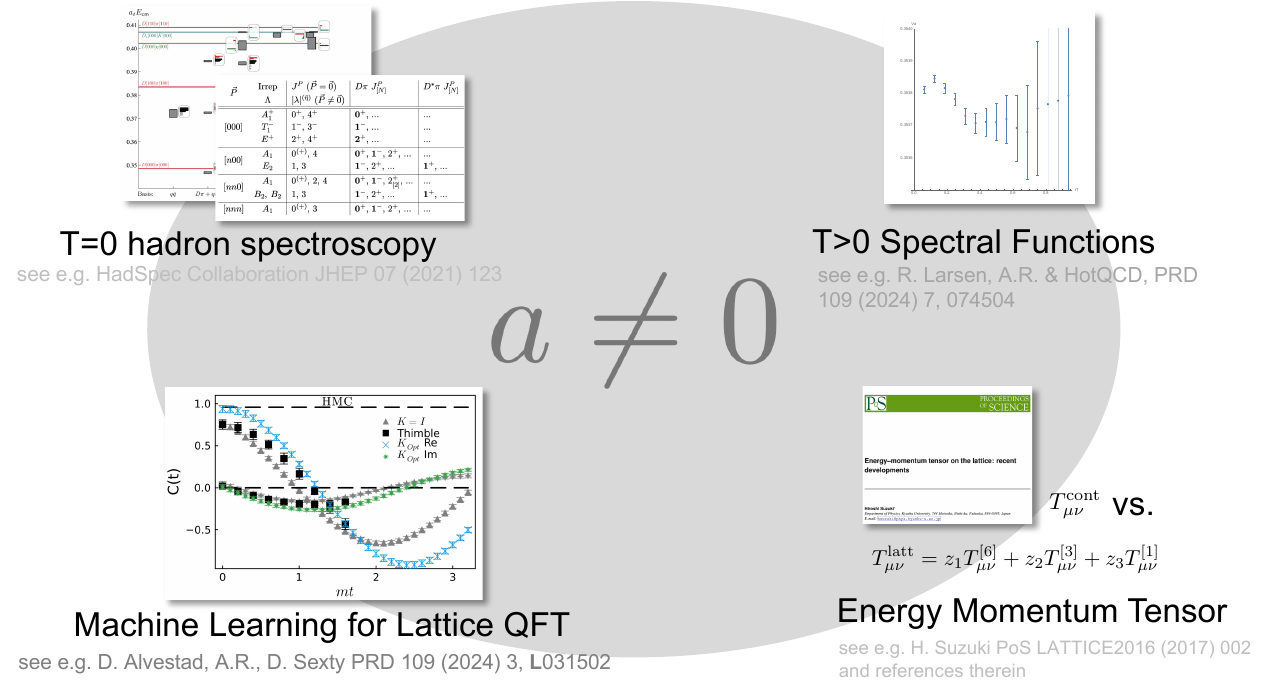}
\caption{Lattice spacing induced symmetry breaking and its effects. See main text for discussion.}
\label{fig:motivation}
\end{figure}

Symmetry breaking induced by the finite lattice spacing $a$, inherent in conventional formulations of lattice field theory, adversely affects the physics interpretation of lattice simulations. In \cref{fig:motivation} we illustrate several concrete examples. In the absence of rotational symmetry, the angular momentum quantum numbers used to label hadron states in the continuum lose their meaning and mixing among physical states ensues (see e.g.~\cite{Gayer:2021xzv}). Lack of translation symmetry is known to lead to unphysical contributions to the energy momentum tensor on the lattice, whose renormalization hence becomes more involved (see e.g.~\cite{Suzuki:2016ytc}). When improved actions are deployed, one benefits from a more rapid convergence towards the continuum limit, but at the cost that these actions may not respect reflection-positivity. In turn the analysis of hadronic spectral functions suffers from the occurrence of positivity violations (see e.g.~\cite{Bala:2021fkm}). Last but not least, the use of machine learning approaches to lattice field theory (see e.g. \cite{Alvestad:2023jgl} and \cite{Sexty:2024lat24} in these proceedings) relies on training data, often informed by continuum physics. If the simulation, on which the machine learning approach is deployed, does not preserve the symmetries present in the training data the mismatch may affect the reliability of the analysis.

In this contribution we present recent work  \cite{Rothkopf:2024hxi}  that realizes a discretization of classical (scalar) lattice field theory, while retaining space-time symmetries. It relies on a novel reparameterization invariant classical action, whose construction is inspired by the work-line formalism of the general theory of relativity. Its key ingredients are dynamical coordinate maps, which evolve together with the field. Deploying a discretization scheme that mimics exactly integration by parts we maintain Noether's theorem in the discrete setting. Using classical scalar wave propagation in $(1+1)d$ as proof-of-principle, we show that the Noether charge remains exactly conserved and also find that the coordinate maps adapt to the dynamics of the field, realizing automatic mesh refinement.

\section{From the world-line formalism to a new action for classical scalar fields}

In the following we will construct a novel action for scalar field theory, borrowing from the formalism of general relativity (GR). We do not consider curved spacetime, $G_{\mu\nu}=\eta_{\mu\nu}={\rm diag}[1,-1,-1,-1]$ but adapt the mathematical framework of reparameterization invariant actions. 

\subsection{Point particle motion in the world-line formalism}

Before we turn to fields, let us first discuss point particle motion in the world-line formalism in GR. In contrast to the non-relativistic treatment of particle motion in terms of a trajectory, i.e.\ a function describing spatial position $\vec{x}(t)$ that depends on the parameter $t$, the world-line formalism requires us to treat space and time on equal footing. The path a single particle traces out as it evolves in spacetime is a one-dimensional manifold and thus can be parameterized by a single number, often denoted by $\gamma\in[0,1]$, christened the world-line parameter. At each point along the path, the particle's spacetime coordinates are represented by dynamical coordinate maps $t(\gamma)$ and $\vec{x}(\gamma)$ which depend on the abstract parameter $\gamma$.

In GR the particle world-line describes a geodesic, the shortest path in a given space-time. Such a geodesic may be obtained from a variational principle, based on the geodesic action
\begin{align}
    S_{\rm geo} = \int \, d\gamma\,(-mc)\Big\{  \sqrt{ \Big(G_{00}+\frac{V(\vec{x})}{2mc^2}\Big)\frac{d X^0}{d\gamma} \frac{dX^0}{d\gamma}+ G_{ii}\frac{d X^i}{d\gamma} \frac{dX^i}{d\gamma} } \Big\},\label{eq:geodesicincl}
\end{align}
where $m$ denotes the mass and $V$ the potential underlying the forces acting on the point particle. This action is but the generalization of path length to an arbitrary geometry described by the metric $G$. When taking the non-relativistic limit, involving slow speeds $v/c\ll 1$ and potential energy smaller than rest energy $V/2mc^2\ll1$ one arrives at the standard action 
\begin{align}
    S_{\rm nr} \overset{\rm from \; \cref{eq:geodesicincl}}{=} \int \, dt\, \Big\{-mc^2 +  \frac{1}{2} m (\dot{\vec{x}}(t))^2 - V\big(\vec{x}(t)\big) \Big\},\label{eq:nonrelacderiv}
\end{align}
with one extra contribution, the constant rest mass term. Note the factor $1/2$ in the kinetic energy, arising from the expansion of the square root in \cref{eq:geodesicincl}. The term $(mc)$ denotes a novel scale, at which particle motion through time becomes inseparable from motion through space. 

As we have showed in a previous study \cite{Rothkopf:2023ljz,Rothkopf:2023vki}, the geodesic action offers the opportunity to discretize in the world-line parameter $\gamma$ instead of time $t$. In turn the coordinate maps remain continuous even in the discrete setting. We were able to establish a Noether theorem for the point particle motion and showed that its energy is exactly preserved at its continuum value. Furthermore we found that $t(\gamma)$ as dynamical coordinate map adjust to the dynamics of the particle, leading to a varying time resolution along the world line, i.e.\ one-dimensional automatic mesh refinement.

\subsection{The Stavanger-Cape Town-Link\"oping (SCL) action for scalar fields}

\begin{figure}
\centering
\includegraphics[scale=0.25]{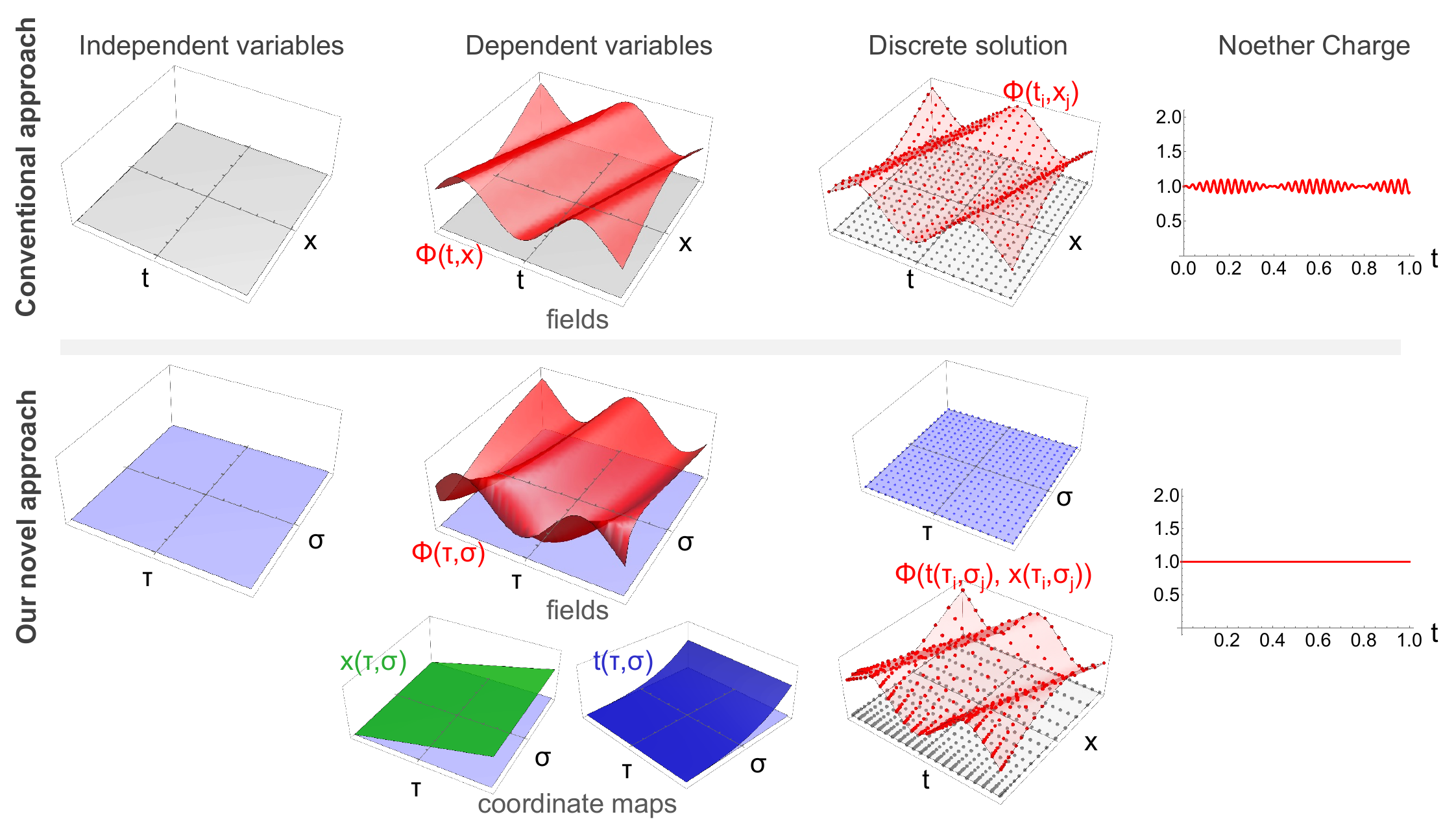}
\caption{Comparison of the (top) conventional formulation of field theory where fields $\phi(t,x)$ lives in space and time and discretization breaks spacetime symmetries. (bottom) Sketch of our novel formalism with dynamical coordinate maps $t(\tau,\sigma)$ and $x(\tau,\sigma)$ in addition to the fields $\phi(\tau,\sigma)$. All degreed of freedom live in an abstract parameter space, whose discretization leaves spacetime symmetries intact (from \cite{Rothkopf:2024hxi}). }\label{fig:formalism}
\end{figure}

Inspired by the world-line formalism we set out to derive a novel reparameterization invariant action for scalar fields. As shown in the lower half of \cref{fig:formalism}, in such a formalism spacetime coordinates are represented by dynamical coordinate maps $t(\tau,\sigma)$ and $x(\tau,\sigma)$, which together with the field $\phi(\tau,\sigma)$ evolve in an abstract space spanned by the parameters $(\tau,\sigma)$. The former denotes temporal evolution, the latter spatial evolution. Just as discretization of the world-line parameter did not affect the space-time symmetry properties the goal here too is to discretize in $(\tau,\sigma)$ instead of $(t,x)$ in order to retain the continuum symmetries.

Let us briefly review the key steps in the derivation, which is discussed in detail in \cite{Rothkopf:2024hxi}. Starting from the conventional reparameterization invariant action, we consider the possibility that this action itself is the low energy limit of another more general action. Taking inspiration from the world-line formalism, we reverse engineer a more general action, starting by including a constant $T$
\begin{align}
    S=\int d^{(d+1)}X \, \sqrt{-{\rm det}[G]} \Big\{ -T + \frac{1}{2}\Big(G^{\mu\nu} \partial_\mu\phi(X)\partial_\nu\phi(X) + V(\phi) \Big)\Big\},\label{eq:stdFTactionpconst}
\end{align}
similar to the term $(mc)$ in \cref{eq:geodesicincl}. Let us move $T$ to the front after which the action appears to represent the lowest order in an expansion of powers of $\kappa={\rm action\, density}/T$
\begin{align}
 {\cal S}_{\rm BVP}&\equiv \int d^{(d+1)}X \, \sqrt{-{\rm det}[G]} \big(-T\big) \Big\{ 1 - \frac{1}{2T}\Big(G^{\mu\nu} \partial_\mu\phi(X)\partial_\nu\phi(X) + V(\phi) \Big)\Big\} + {\cal O}(\kappa^2),\\
    &\approx\int d^{(d+1)}X \, \sqrt{-{\rm det}[G]} \big(-T\big) \sqrt{ 1 - \frac{1}{T}\Big(G^{\mu\nu} \partial_\mu\phi(X)\partial_\nu\phi(X) + V(\phi) \Big)}.  \label{eq:stdFTactionbacksqrt}
\end{align}
Reverse engineering the action by undoing the assumed expansion, we arrive in \cref{eq:stdFTactionbacksqrt} at an action ${\cal S}_{\rm BVP}$ that contains a square root term. This action is still formulated in terms of spacetime coordinates. The next decisive step is to elevate the coordinates to dynamical coordinate maps $X^\mu\to X^\mu(\Sigma)$ dependent on the abstract parameters $\Sigma^a=(\tau,\vec{\sigma})^a$. Changing the integration variables $X\to\Sigma$ introduces the determinant of the Jacobian $|{\rm det}(J)|$, which together with the metric term $\sqrt{-{\rm det}[G]}$ combines into the determinant of the so-called induced metric $g_{ab}\equiv G_{\mu\nu} J^\mu_a J^\nu_b$ describing the geometry of the space of parameters
\begin{align}
    {\cal S}_{\rm BVP}&=\int d^{(d+1)}\Sigma \, \big(-T\big) \sqrt{ -{\rm det}[g] + \frac{1}{T}\Big(G^{\mu\nu} \partial_\mu\phi(\Sigma)\partial_\nu\phi(\Sigma) + V(\phi) \Big){\rm det}[g]}.  \label{eq:stdFTactionbacksqrtmod2}
\end{align}
There are still references to the old metric and spacetime derivatives visible under the square root, which we can convert by introducing the adjugate of the induced metric  ${\rm adj}[g]=g^{-1}{\rm det}[g]$ to obtain the final form of the Stavanger-Cape Town-Link\"oping (SCL) action 
\begin{align}
{\cal S}_{\rm BVP}=\int d^{(d+1)}\Sigma \, \big(-T\big) \sqrt{ \Big(\frac{1}{T}V(\phi)-1\Big){\rm det}[g] + \frac{1}{T} \partial_a\phi(\Sigma)\partial_b\phi(\Sigma) {\rm adj}[g]_{ab}}.\label{eq:novelaction}
\end{align}
Thus we achieved the goal of introducing dynamical coordinate maps along with the field. These maps are the key that allows us to discretize the action in the abstract parameters $(\tau,\vec{\sigma})$ and they implement automatic mesh refinement as we show in \cref{sec:proof}. Note that $T$ represents the scale where dynamics of the field and that of coordinate maps become inseparable.

Classical dynamics are fully specified by the critical point of the classical action. Since the critical point does not change when applying a monotonous function to the integrand of the action, we go over to the following expression devoid of the square root
\begin{align}
 {\cal E}_{\rm BVP}&=\int d^{(d+1)}\Sigma \; \frac{1}{2} \Big\{ \Big(\frac{1}{T}V(\phi)-1\Big){\rm det}[g] + \frac{1}{T} \partial_a\phi(\Sigma)\partial_b\phi(\Sigma) {\rm adj}[g]_{ab}\Big\}. \label{eq:novelactionE}
\end{align}


\section{Summation-by-parts finite difference discretization of the novel action}

Our goal is discretize the action of \cref{eq:novelactionE} in the abstract parameters $(\tau,\vec{\sigma})$ to leave the coordinate maps continuous. This will allow us to retain the continuum space-time symmetries of the system. In addition in order for symmetries to be related to conserved quantities, Noether's theorem must hold. To derive it from the system action, integration by parts is a crucial ingredient. Thus we deploy a finite difference scheme that mimics integration by parts exactly in the discrete setting, fulfilling the so-called summation by part (SBP) property (for reviews see \cite{svard2014review,fernandez2014review,lundquist2014sbp}).

Let us discuss the SBP approach in a simple one-dimensional setting, with functions $u(\tau)$ and $v(\tau)$, which we discretize as ${\bf u}$ and ${\bf v}$ on an equidistant grid with $N$ points with $\Delta\tau=(\tau_f-\tau_i)/(N-1)$. Since integration by parts connects integration and differentiation, we must first select a quadrature rule $\int_{\tau_i}^{\tau_f} d\tau\, u(\tau)\, v(\tau) \approx {\bf u}^t \,{\mathds H} \,{\bf v} = ({\bf u},{\bf v})$ via the diagonal positive definite matrix $\mathds H$. The associated SBP finite difference operator $\mathds{D}$ is constructed as
\begin{align}
&\mathds{D}=\mathds{H}^{-1}\,\mathds{Q}, \quad \mathds{Q}+\mathds{Q}^T=\mathds{E}_N-\mathds{E}_1={\rm diag}[-1,0,\ldots,0,1],\label{eq:SBP}
\end{align}
where $\mathds{Q}$ encodes the stencil structure of the finite difference. The matrices $\mathds{E}_i$ contain zero everywhere, except at the i-th diagonal entry. It is the relation on the right of \cref{eq:SBP} that establishes the summation-by-parts nature of $\mathds{D}$, i.e. it guarantees the correct boundary treatment in the inner product $({\mathds{D} \bf u})^T \,{\mathds H} \,{\bf v} = - {\bf u}^T \,{\mathds H} \,{ \mathds{D}\bf v} + {\bf u}_N{\bf v}_N - {\bf u}_0{\bf v}_0$. The lowest order SBP operator \texttt{SBP121} reads
\begin{align}
\mathds{H}^{[121]}=\Delta t\left[\begin{array}{cccc} \frac{1}{2}&&&\\ &1&&\\&&1&\\&&&\frac{1}{2}\end{array}\right],\quad \mathds{D}^{[121]}=\frac{1}{2\Delta t}\left[ \begin{array}{ccccc} -2 &2 & & &\\ -1& 0& 1& &\\ & &\ddots && \\ &&-1&0&1\\ &&&-2&2 \end{array} \right].
\end{align}
Note that it consists of the forward and backward derivative stencil on the boundaries, as well as the central symmetric stencil in the interior (this operator has been investigated in the context of point particle quantum mechanics in \cite{Kim:2024bcx}).

\begin{figure}
\centering
\includegraphics[scale=0.23]{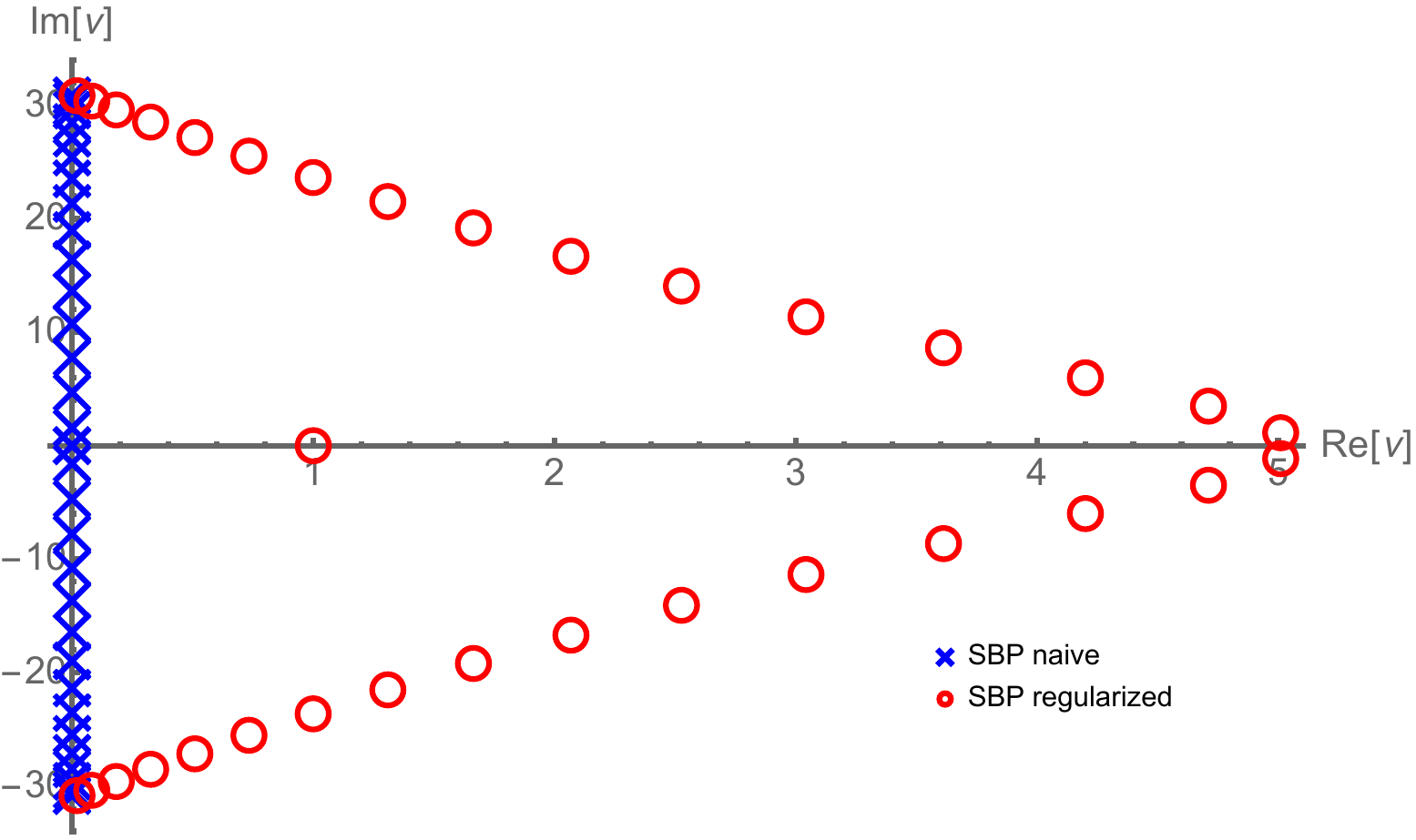}
\caption{(blue) Spectrum of the naive \texttt{SBP121} finite difference operator on a $N=32$ grid. Note that two zero modes, one physical, one unphysical occur. (red) Spectrum of the regularized SBP operator is devoid of zero modes and the physical constant function is represented by the eigenvalue unity (from \cite{Rothkopf:2022zfb}). }\label{fig:SBPspec}
\end{figure}

As is well known in the lattice community, a central symmetric finite difference operator suffers from the infamous doubler problem. The strategy proposed by Wilson in the case of fermion fields is not applicable here, as it requires modifying the finite difference with a complex term. Here our operator acts only on purely real functions. Instead we developed an alternative regularization strategy \cite{Rothkopf:2022zfb} which is inspired by the weak treatment of boundary conditions in modern approaches to partial differential equations. Our strategy is to incorporate penalty terms into the finite difference operator, which penalize deviation from the boundary data. I.e. we define a new regularized SBP operator $\bar{\mathds D}{\bf u}=\mathds D{\bf u} + {\mathds H}^{-1}\mathds{E}_1({\bf u}-{\bf u}_1)$,
where the array ${\bf u}_1$ contains the boundary data in its first entry. This modification contains two contributions. The first, proportional to ${\bf u}$ can be readily absorbed into the original structure of $\mathds D$. However to accommodate the constant shift related to ${\bf u}_1$ we must introduce affine coordinates, i.e. to construct $\bar{\mathds D}$, the original $\mathds D$ is extended by one row and one column, the value unity placed in the new bottom right corner. The shift then enters in the newly added column. As shown in \cref{fig:SBPspec} the unphysical zero mode of the original $\mathds D$ operator (blue cross at the origin) is lifted and the original physical zero mode is turned into the single eigenvector with a purely real eigenvalue one (see red circles).

With the properly regularized SBP operator at hand (a detailed description of how to generalize the above construction to multiple dimensions is provided in \cite{Rothkopf:2024hxi}) we can proceed to discretize our novel action. Note that for each degree of freedom in the action, the regularization of the SBP operator makes reference to its specific boundary data. I.e.\ we introduce differently regularized SBP operators, denoting them by an additional superscript e.g.\ $\partial_a \phi \approx {\bar{\mathds{D}}_a^\phi} {\bm \phi}$ or $\partial_a X^\mu \approx {\bar{\mathds{D}}_a^\mu} {\bm X}^\mu$ (in the latter case no summation is implied). Introducing the vector ${\bf h}$ with entries $h_i={\mathds H}_{ii}$ we obtain
\begin{align}
\hspace{-0.25cm} \mathds{E}_{\rm BVP}&[{\bm X}^\mu,\bar{\mathds{D}}^\mu_a {\bm X}^\mu,{\bm \phi}, \bar{\mathds{D}}^\phi_a{\bm \phi}]=\frac{1}{2}\Big\{ \Big(\frac{1}{T}V({\bm \phi})-1\Big)\circ{\rm det}[{\bm g}] +\frac{1}{T} (\bar{\mathds{D}}^\phi_a{\bm \phi}) \circ  (\bar{\mathds{D}}^\phi_b{\bm \phi})\circ{\rm adj}[{\bm g}]_{ab}\Big\}^T {\bf h}.\label{eq:discrEBVP}
\end{align}
All information about the coordinate maps is encoded in the induced metric $ {\bm g}_{ab}=G_{\mu\nu} (\bar{\mathds{D}}^\mu_a {\bm X}^\mu)\circ (\bar{\mathds{D}}^\nu_b {\bm X}^\nu)$,
where $\circ$ denotes element-wise multiplication of neighboring discrete arrays. Note that \textit{even after discretization}, the entries of ${\bm g}_{ab}$ \textit{remain manifestly invariant under Poincar\'e transformations}.

Since integration by parts is exactly mimicked in the discrete setting, the derivation of Noether's theorem remains intact. All we need to do is to replace derivatives with finite differences and continuum functions with their discretized counterparts in the continuum Noether current.

In order to use \cref{eq:discrEBVP} to solve for the causal evolution of the scalar field and coordinate maps, we will have to cast it in the form appropriate for an initial boundary value problem. This entails going over to a setting in which the degrees of freedom are doubled and which amounts to the classical limit of the Schwinger-Keldysh closed time path formalism. In addition, the initial and boundary data necessary for a unique solution of the dynamics of the system must be provided, which is achieved by introducing additional Lagrange multipliers in \cref{eq:discrEBVP}. A discussion of this construction goes beyond the scope of this contribution, an extended exposition is provided in \cite{Rothkopf:2024hxi}.

\section{Proof-of-principle: classical scalar wave propagation in $(1+1)d$}
\label{sec:proof}

\begin{figure}
\centering
\includegraphics[scale=0.25]{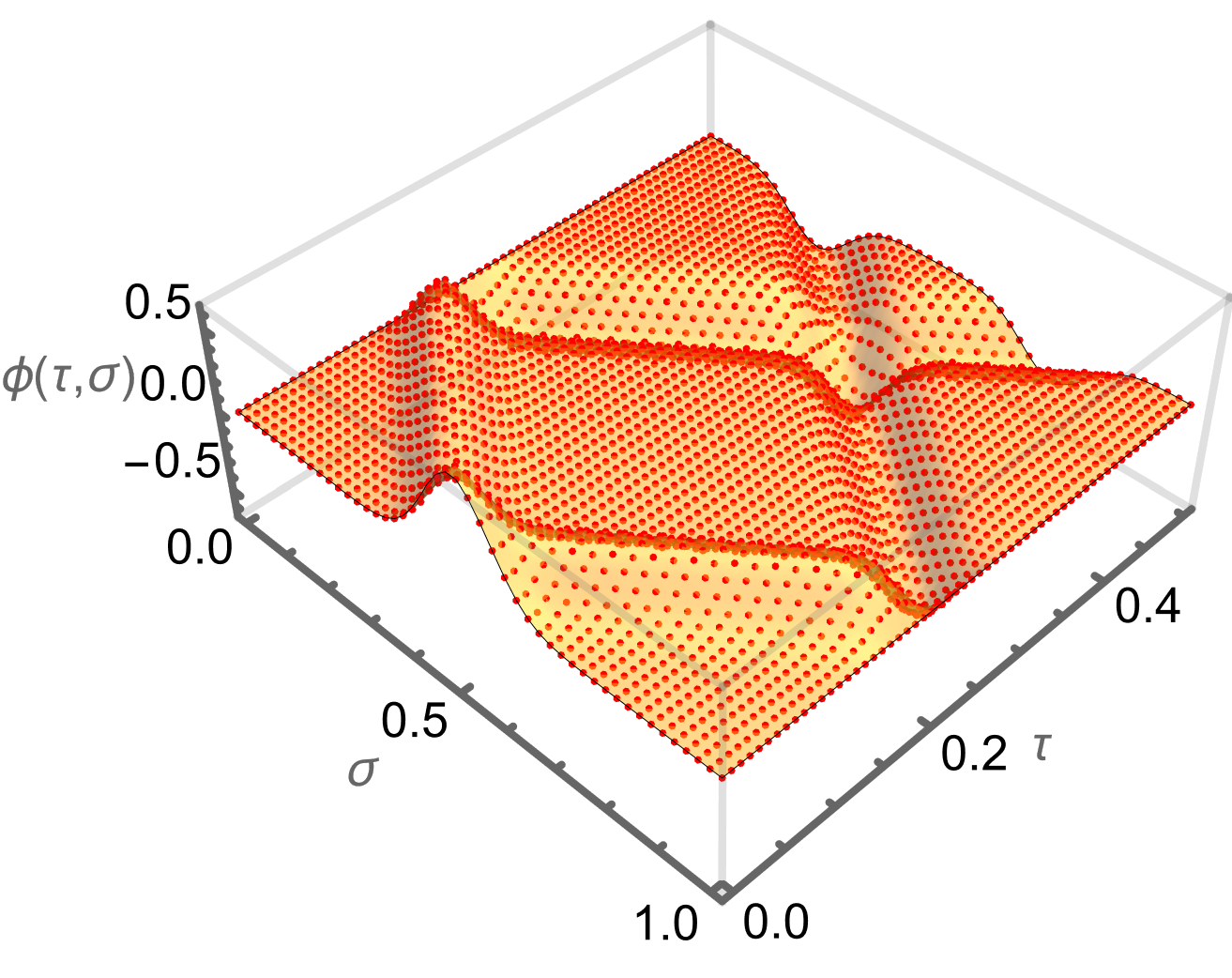}
\includegraphics[scale=0.25]{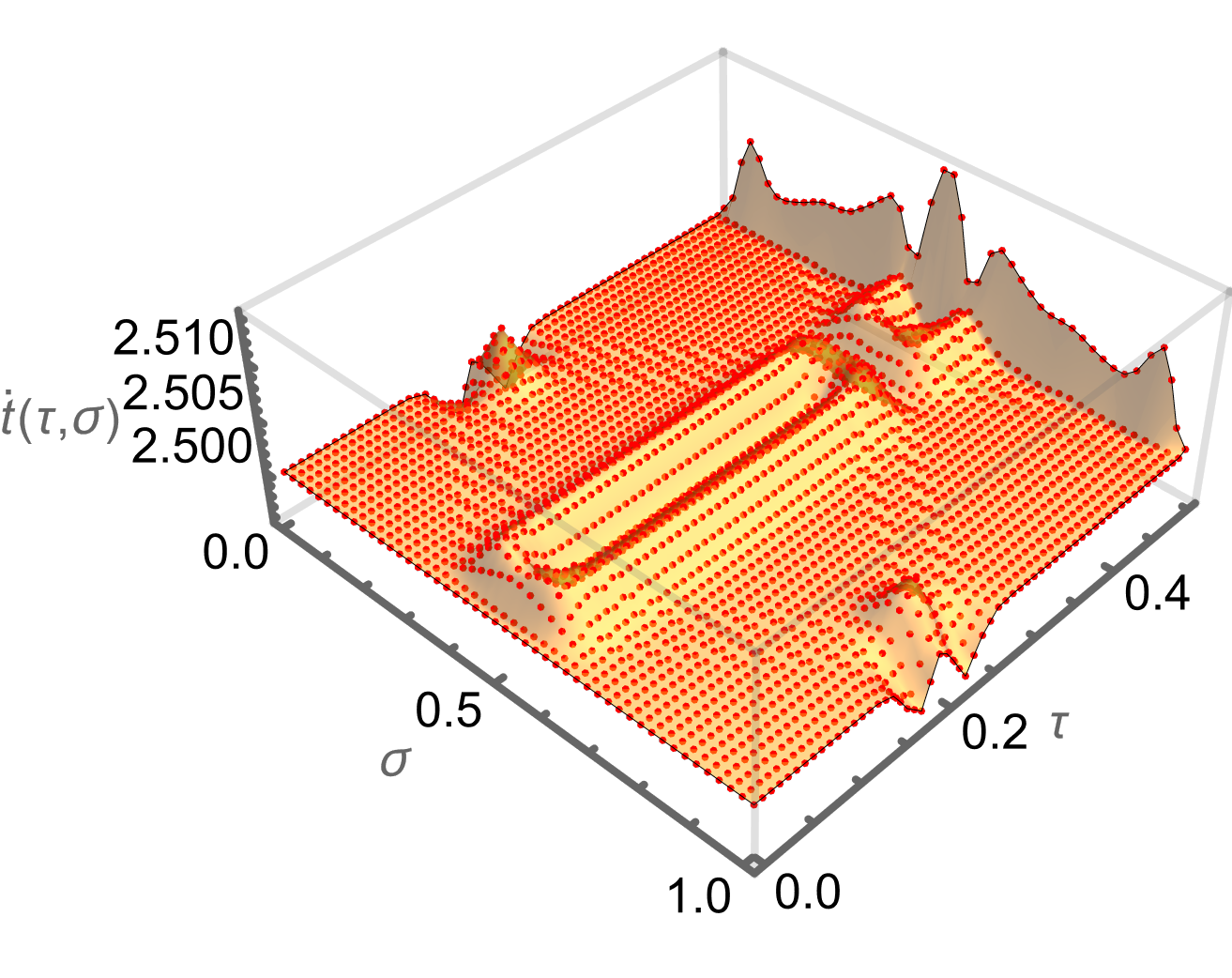}
\caption{(left) Scalar wave propagation from the critical point of the IBVP action based on \cref{eq:bvpEdisc} on a $N_\sigma=48$ and $N_\tau=60$  grid. (right) $\tau$ derivative  $\dot t_{\rm cl}[\tau,\sigma]$ showing adaptive mesh refinement (from \cite{Rothkopf:2024hxi}). }\label{fig:clsol}
\end{figure}

Here we present as proof of principle an application of our approach to scalar wave propagation in $(1+1)$ dimensions. The two abstract parameters in our action are $\tau$ and $\sigma$ and we choose (in absence of an estimate of T in nature) arbitrarily $T=10^4$. We simplify the setting by taking $V=0$ and choose a trivial spatial coordinate map $x[\tau,\sigma]=\sigma$. This leaves us with the action
 \begin{align}
{\cal E}_{\rm BVP}\overset{x=\sigma}{=}\int &d\tau d\sigma \, \frac{1}{2} \Big\{ (\td)^2+\frac{1}{T} \Big( \pd^2((\tp)^2-1) - 2 \pd\pp\td\tp + (\pp)^2(\td^2) \Big) \Big\},\label{eq:novelactionE1p1}
\end{align}
which remains manifestly invariant under global continuous time translations. Discretizing this action on a $N_\tau\times N_\sigma$ grid with appropriately regularized \texttt{SBP121} operators $\mathds{D}_\tau$ and $\mathds{D}_\sigma$  leads to 
\begin{align}
    \nonumber \mathds{E}_{\rm BVP} =& \frac{1}{2}\Big\{ (\bar{\mathds{D}}_\tau^t {\bm t})^2+\frac{1}{T} \Big( (\bar{\mathds{D}}_\tau^\phi {\bm \phi})^2\circ\big( (\mathds{D}_\sigma {\bm t})^2 - 1 \big)\\
&\qquad -2(\bar{\mathds{D}}_\sigma^\phi {\bm \phi})\circ(\bar{\mathds{D}}_\tau^\phi {\bm \phi})\circ (\bar{\mathds{D}}_\tau^t {\bm t})\circ(\mathds{D}_\sigma^t {\bm t})  + (\bar{\mathds{D}}_\sigma^\phi {\bm \phi})^2\circ (\bar{\mathds{D}}_\tau^t {\bm t})^2 \Big)\Big\}^T {\bm h}.\label{eq:bvpEdisc}
    \end{align}    
After incorporating this action in the doubled degrees of freedom formalism and amending it by Lagrange multipliers for boundary data, we numerically obtain its critical point in terms of ${\bm \phi}_{\rm cl}\approx \phi_{\rm cl}[\tau,\sigma]$ and ${\bf t}_{\rm cl}\approx t_{\rm cl}[\tau,\sigma]$ using the \texttt{IPOPT} library implemented in the \texttt{NMinimize} command of \texttt{Mathematica}. On the left of \cref{fig:clsol} we show the evolution of the field on a $N_\sigma=48$ and $N_\tau=60$ grid initialized by a bump centered in the spatial domain. Two wave packages emerge from the bump, traveling to the boundary of the simulation domain and reflecting from the Dirichlet boundary before merging again with inverted amplitude. 

Let us take a look at the evolution of the dynamical temporal coordinate map $t[\tau,\sigma]$ whose $\tau$ derivative we plot on the right of \cref{fig:clsol}. A larger value denotes a coarser time resolution, a smaller value a finer time resolution. We find that the temporal resolution adapts to the dynamics of the field. When the wave packages recede from the interior of the simulation domain the resolution is coarsened, when they violently reflect from the boundary the resolution is self-consistently adjusted. This amounts to \textit{automatic adaptive mesh refinement}.

\begin{figure}
\centering
\includegraphics[scale=0.22]{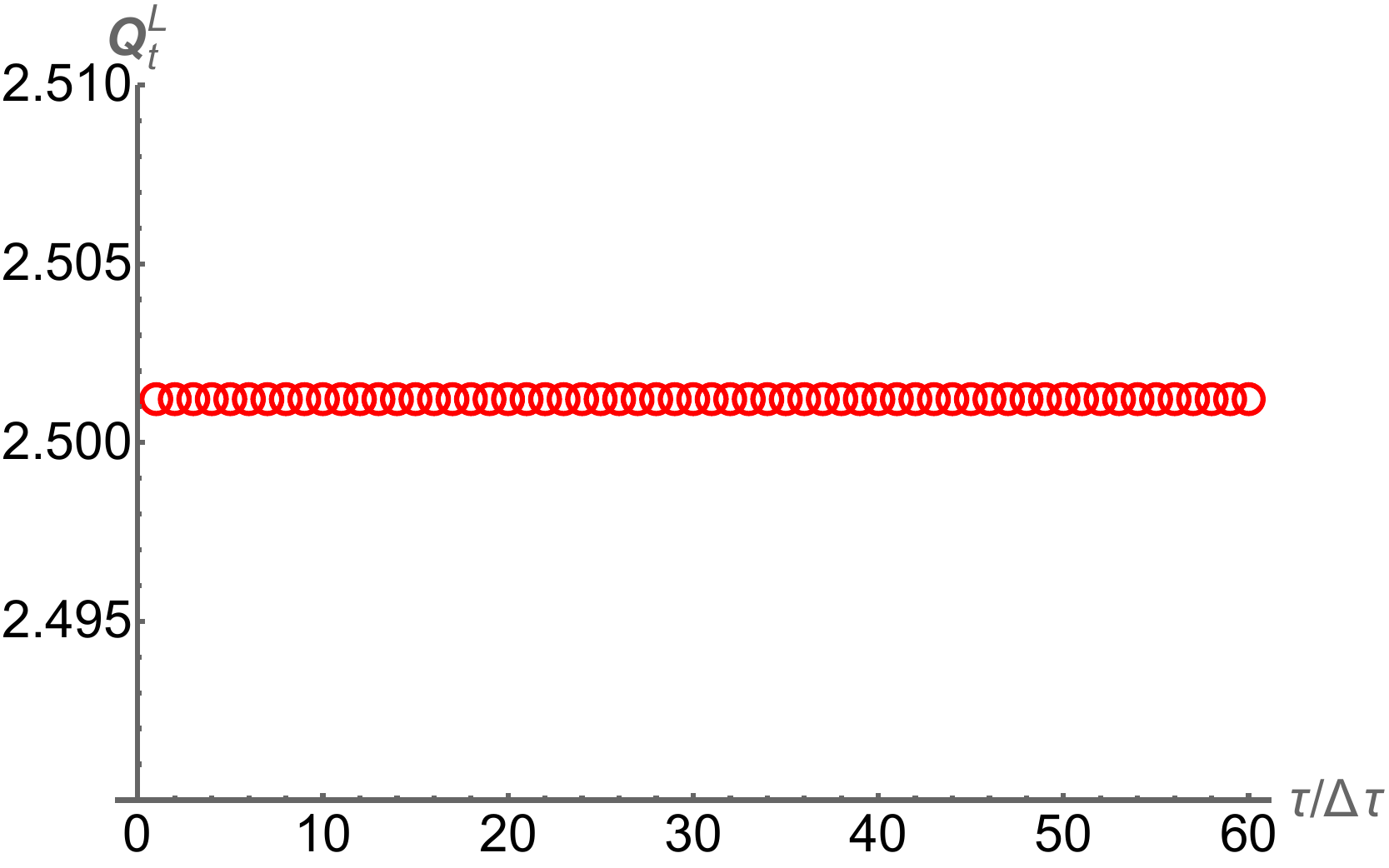}
\caption{Exact conservation of the Noether charge associated with time translation symmetry (from \cite{Rothkopf:2024hxi}).}\label{fig:Noether}
\end{figure}

Let us investigate the expression for the Noether charge associated with the time translation symmetry. For completeness we provide the full expression
\begin{align}
   \nonumber{\bm Q}_t= &\mathds{H}_\sigma \underbracket{\Big\{ ( \mathds{D}_\tau{\bm t}) + \frac{1}{T} \Big( (\mathds{D}_\sigma {\bm \phi})^2\circ(\mathds{D}_\tau{\bm t)}- (\mathds{D}_\tau {\bm \phi})\circ(\mathds{D}_\sigma {\bm \phi})\circ(\mathds{D}_\sigma {\bm t})\Big)\Big\}}_{{\bm J}^0\in {\mathbb{R}^{N_\tau\times N_\sigma}}}\\
   & \hspace{5.4cm}+\underbracket{\Big\{ ({\bm h}_\sigma^T \tilde{\bm \lambda}^t)  \mathfrak{d}^\tau[0]+ ( {\bm h}_\sigma^T \tilde{\bm \gamma}^t ) \mathfrak{d}^\tau[N_\tau]\Big\}}_{\rm Lagr.\, mult.\, contrib.},\label{eq:dsicrNoetherl}
\end{align}
where additional Lagrange multiplier contributions are explicitly shown (for details see \cite{Rothkopf:2024hxi}). These contributions are a reminder that boundary data must be fixed explicitly in the action. As shown in \cref{fig:Noether} we find that this \textit{Noether charge is exactly preserved at its initial value} down to the numerical precision used to obtain the critical point. This conservation goes beyond that offered by symplectic approaches, which only conserve energy on average. It is the exact conservation of the Noether charge that underlies the automatic mesh refinement observed in $t[\tau,\sigma]$, forcing the temporal map to adapt, in order to leave \cref{eq:dsicrNoetherl} constant.

\section{Summary}
We have presented a novel reparameterization invariant action for scalar fields \cite{Rothkopf:2024hxi}, which incorporates dynamical coordinate maps. These maps together with the field propagate in an abstract parameter space. Discretizing in these parameters leaves the coordinate maps continuous and the spacetime symmetries intact. Deploying SBP operators allows us to retain Noether's theorem. To avoid doublers, we rely on an alternative to the Wilson term \cite{Rothkopf:2022zfb}, which exploits boundary data and is applicable also to purely real functions. Using classical scalar wave propagation as proof of principle, we show that the Noether charge remains exactly conserved, which in turn forces the time mapping to adapt to the dynamics, leading to automatic adaptive mesh refinement.\\

\vspace{-0.6cm}
\acknowledgments
A.\ R.\ and W.\ A.\ H.\ acknowledge support by the ERASMUS+ project 2023-1-NO01-KA171-HED-000132068. W.\ A.\ H.\ thanks the South African National Research Foundation and the SA-CERN Collaboration for financial support. J.\ N.\ was supported by the Swedish Research Council grant nr. 2021-05484 and the University of Johannesburg.

\bibliographystyle{JHEP}
\bibliography{Lattice2024}

\end{document}